\documentclass{article}
\usepackage{amsmath,amsthm,amssymb}
\usepackage{multirow}
\usepackage{authblk}
\usepackage{url}
\usepackage{graphicx}
\usepackage{float}
\usepackage{subfig}
\floatstyle{boxed}
\restylefloat{figure}
\usepackage[utf8x]{inputenc}
\usepackage{hyperref} 
\usepackage{enumerate}
\usepackage{bm}
\title{Return probability and $k$-step measures}
\author{Nicholas Dronen}
\affil{Department of Computer Science\\
    University of Colorado at Boulder\\
    dronen@colorado.edu}
\affil{Pearson Knowledge Technologies\\
    Boulder, Colorado, USA}
\author{Qin Lv}
\affil{Department of Computer Science\\
    University of Colorado at Boulder\\
    qin.lv@colorado.edu}

\begin{document}
\date{\today}
\maketitle

\begin{abstract}
The notion of \emph{return probability} -- explored most famously by
George P\'{o}lya on $d$-dimensional lattices -- has potential as a 
measure for the analysis of networks.  We present an efficient method 
for finding return probability distributions for connected undirected 
graphs.  We argue that return probability has 
the same discriminatory power as existing $k$-step measures -- in 
particular, beta centrality (with negative $\beta$), the 
graph-theoretical power index (GPI), and subgraph centrality.
We compare the running time of our algorithm to beta centrality and
subgraph centrality and find that it is significantly faster.  When return
probability is used to measure the same phenomena as beta centrality,
it runs in linear time -- $O(n+m)$, where $n$ and $m$ are the number 
of nodes and edges, respectively -- which takes much less time than 
either the matrix inversion or the sequence of matrix multiplications 
required for calculating the exact or approximate forms of beta 
centrality, respectively.  We call this form of return probability
the \emph{P\'{o}lya power index} (PPI).  Computing subgraph centrality
requires an expensive eigendecomposition of the adjacency matrix;
return probability runs in half the time of the eigendecomposition
on a 2000-node network.  These performance improvements are important
because computationally efficient measures are necessary in order
to analyze large networks.
\end{abstract}

\section{Introduction}

The probability that a random walk on a graph returns to the node
where it began -- the \emph{probability of returning to the origin} or
simply \emph{return probability} -- is a fairly well-known notion in the
literature of random walks.  Research in this area originally concentrated
on return probability on infinite regular graphs.  In his seminal 
work~\cite{Polya1921}, P\'{o}lya proved that a random walk on an infinite 
1- or 2-dimensional lattice returns to the origin with probability $p=1$, 
but when $d>2$, $p<1$.  Methods for determining the value of $p$ for
3-dimensional lattices were subsequently discovered~\cite{Watson1939,
McCrea1940, Domb1954, Glasser1977}.  P\'{o}lya's theorem has also been 
applied to electrical networks by Doyle and Snell~\cite{Doyle2000}.  
Return probability continues to be explored in contemporary research, 
although the venue has shifted from graphs of fixed degree to random 
graphs~\cite{Grimmett2010, Martin2010, Heicklen2005} and to spectral 
methods~\cite{Sobieczky2010}.

There is a class of measures which compute some value for a node
$i$ based on paths up to length $n$ originating at $i$.  Some
representative members of this class are degree centrality, beta
centrality~\cite{Bonacich1987}, the graph-theoretical power index
(GPI) (see~\cite{Willer2008} for an overview), and subgraph
centrality~\cite{Estrada2005}.  These measures have been called
``$n$-path centralities'' \cite{Borgatti2006}.  This term is
problematic.  First, a measure is only a centrality when it satisfies
certain requirements, such as those proposed in~\cite{Ruhnau20003}.
Beta centrality with negative $\beta$ and the GPI, however, are not
centralities.  ``Path'' is infelicitous, too, because each
measure pays attention to different entities.  Beta centrality is
based on walks, the GPI counts disjoint paths, and subgraph centrality
is derived from closed walks.  We propose to refer to them
instead as ``$k$-step measures.''

Return probability is a $k$-step measure as well, and it has
a few virtues that distinguish it from the others: (1) being a probability,
it is always in the range $[0, 1]$ and requires no normalization, so
the return probability of two nodes can always be meaningfully compared,
even when the nodes are in different networks; (2) it permits precise
control of the length of walks over which it is computed; and
(3) it can be computed very efficiently. 

The notation used in this article is mostly conventional.  
We only consider graphs $G = (V, E)$ that
are simple, connected, and undirected.  Let $n = |V|$ be the number
of vertices, and $m = |E|$ be the number of edges.  The length
of some sequence of adjacent vertices -- e.g., a path or a walk --
is denoted by $k$.  Let $\boldsymbol{A} = \boldsymbol{A}(G)$ be the adjacency matrix of $G$,
where $a_{ij} = 1$ if there is an edge between $i$ and $j$ and
$a_{ij} = 0$ otherwise.  Let $\boldsymbol{P} = \boldsymbol{P}(G)$ be the transition probability
matrix of $G$, where $p_{ij} = 1/deg(i)$ if $i$ and $j$ are adjacent,
and $p_{ij}=0$ otherwise.  We denote the probability of an event $X$ by $\mathbb{P}(X)$.
We occasionally diverge
from convention.  In \autoref{Algorithm} we use $\boldsymbol{A}^{(k)}$ -- with parens that distinguish it from the usual $\boldsymbol{A}^{k}$ -- to
indicate a kind of $k$-th power of $\boldsymbol{A}$ that is essential for computing
return probability.  And in \autoref{Power measures} we abuse the
$\gg$ and $\ll$ symbols to compress our visual comparison of the
power-related $k$-step measures.

Beta centrality and the GPI are measures of exclusionary power\footnote{
Alter-based centrality \cite{RePEc:boc:bocode:s457186} is 
notably similar to return probability with $k=2$.  In its negative
mode, it can, like return probability, be used in lieu of beta
centrality with negative $\beta$.}.
They identify the relative power of nodes in a network by their
ability to exclude their neighbors from some valuable interaction.
Beta centrality of node $i$ is the $i$-th component of the vector: 
\begin{equation*}
    C(\beta) = \sum_{k=0}^\infty \beta^{k-1} \boldsymbol{A}^k 1
\end{equation*}
According to Bonacich~\cite{Bonacich1987}, the 
``sign of $\beta$ corresponds exactly
to the distinction \ldots between positive and negative exchange
systems'' and its magnitude ``affects the degree to which distant
ties are taken into account''.  In this article,
we are only interested in beta centrality with negative values of 
$\beta$.  The GPI is defined as:
\begin{equation*}
GPI_{i}(e) = \sum_{k=1}^{g} (-1)^{(k-1)} m_{ik}
\end{equation*}
where $g$ is the diameter of the network, $m_{ik}$ is the number
of non-intersecting paths of length $k$ originating at node $i$,
and $e$ is the number of exchange opportunities that node $i$ has
in any round\footnote{The GPI has undergone many changes since
its inception.  This equation for the GPI appeared early in the
tussle of theories competing in the exchange network literature in
the '80s and '90s.  Improved methods, results of which we use later in this paper, focus on the probability of
a node being excluded in a round of exchanges.}.  Finally, subgraph
centrality~\cite{Estrada2005} -- a measure of the number of subgraphs
in which a node participates -- is defined as:
\begin{equation*}
    C_s(i) = \sum_{k=0}^\infty \frac{(\boldsymbol{A}^k)_{ii}}{k!}
\end{equation*}

Later we discuss in detail other connections among beta centrality,
the GPI, subgraph centrality, and return probability.  For now it
suffices to note a few characteristics shared just by beta centrality
and subgraph centrality, and to situate return probability in
relation to them.  First, beta centrality and subgraph centrality are formally
expressed as involving increasing powers of the adjacency matrix
$\boldsymbol{A}$.  Return probability is expressed in a similar way (although
in practice we use a stochastic matrix), but each power of $\boldsymbol{A}$ must
be modified before the subsequent power can be computed.  
Second, beta
centrality and subgraph centrality are expressed as infinite sums.
Since cumulative return probability converges to 1 as the walk
length $k \rightarrow \infty$, the return probability for each node
is the same in the limit, which is not informative.  Instead we
find a \emph{distribution} of return probabilities over walks of
length $1, \ldots, k$.  Finally, beta and subgraph centrality are
scalar, assigning a single real value to a node.  In contrast, a
distribution of return probabilities is a sequence of real values.
To reduce a distribution of return probabilities for a node to a
scalar value, we take either the return probability or the cumulative
return probability at some chosen $k$.  This allows us to compare
return probability to other measures.

The purpose of this article is to show that return probability is
equivalent to these three $k$-step measures,
and that it can be computed more efficiently, much more so in some cases.
The rest of this article is organized as follows.
In Section \ref{Computing Return Probability} we propose and validate a method for finding return
probabilities.  It is based on a particular kind of walk -- the
\emph{self-absorbing walk} -- which we use to model the probability
of returning to the origin for the first time.  If return probability
is a useful measure, what does it measure?  We devote Section \ref{Return Probability and Other Measures}
to that question, showing that with $k=2$, return probability is
strongly related to existing power measures, implying that return
probability is at least an approximation of exclusionary power.  We
call this measure the \emph{P\'{o}lya power index} (PPI).  We
also show that return probability with $k>2$ is equivalent to
subgraph centrality.  Finally, we show that return probability is
significantly more efficient to compute than beta centrality and
subgraph centrality.  Section \ref{Discussion} contains further discussion.
Section \ref{Conclusion} concludes.

\section{Computing Return Probability} \label{Computing Return Probability}
\subsection{Algorithm} \label{Algorithm}

Consider a random walk on a graph $G = (V, E)$.  Choose some node
$i \in V$ as the origin and begin to walk.  If we return to $i$,
the walk terminates, and we start a new walk.  To emphasize that
in these walks $i$ becomes a terminating point only after the walk
leaves $i$, we call this a \emph{self-absorbing walk}.  With such
walks, returning to the origin at step $k$ is mutually exclusive
with returning to the origin before step $k$.  Thus the probability
of returning to the origin in a $k$-step walk is related to the 
following two probabilities:
\begin{enumerate}
\item The probability of returning to $i$ at step $k$.
\item The probability of not returning to $i$ at any step $<k$.
\end{enumerate}

For the first probability, let $Next_{i,k}$ be the event of returning
to $i$ at step $k$ on a self-absorbing walk originated from $i$.  
To compute $\mathbb{P}(Next_{i,k})$, we must know the states we can potentially be in on a
walk of length $k-1$.  From there we must count the number of next steps
that are possible from that set of states, taking care to distinguish
those that return to $i$ from those that do not.  Define $Steps_{i,k}$ as the
number of possible next steps from the set of possible states after a walk
of length $k-1$ and define $ReturnSteps_{i,k}$ as the number of possible
next steps that return to $i$ from the same set of possible states.
Then $\mathbb{P}(Next_{i,k})$ is:
\begin{equation}
\label{PNext}
\mathbb{P}(Next_{i,k}) = 
\begin{cases}
    ReturnSteps_{i,k}/Steps_{i,k} & \text{if } Steps_{i,k} > 0 \\
    0 & \text{if } Steps_{i,k} = 0
\end{cases}
\end{equation}

The second of these two probabilities is the complement of the probability
of returning in any step $< k$.  Let $R_{i,k}$ denote the event of
returning to the origin $i$ at step $k$.  Then the probability of not
returning to the origin in $k-1$ steps is: 
\begin{align}
\mathbb{P}(\overline{R_{i,1}} \land \ldots \land \overline{R_{i,k-1}}) &= \mathbb{P}(\overline{(R_{i,1} \lor \ldots \lor R_{i,k-1})}) \nonumber \\
&= 1-(\mathbb{P}(R_{i,1}) \lor \ldots \lor \mathbb{P}(R_{i,k-1})) \nonumber \\
&= 1-(\mathbb{P}(R_{i,1}) + \ldots + \mathbb{P}(R_{i,k-1})) \nonumber \\
&= 1 - \sum_{x=1}^{k-1}\mathbb{P}(R_{i,x}) \label{PCond}
\end{align}

Combining \eqref{PNext} and \eqref{PCond} yields our equation for the
return probability for any node $i \in V$ and any length $k$:
\begin{equation}
\label{PRet}
\mathbb{P}(R_{i,k}) = \left[1 - \sum_{x=1}^{k-1}\mathbb{P}(R_{i,x}) \right] \mathbb{P}(Next_{i,k})
\end{equation}

It is well known that an element $a_{i,j}^{k}$ of the $k$-th power
of $\boldsymbol{A}$ is the number of walks of length $k$ from node $i$ to node
$j$.  A non-zero element $a_{i,i}^{k}$ indicates the number of
closed walks of length $k$ originating at $i$.  If the diagonal of
$\boldsymbol{A}$ and its powers are left undefined, the same process computes
simple paths instead of walks.  However, neither technique counts
self-absorbing walks.  The first one fails to terminate a walk once
it returns to the origin, causing it to be counted more than once; 
and because the diagonal is all zeros in the second one, it disallows 
returning to the origin altogether.  To count self-absorbing walks, our
computation must permit a walk to return to the origin and must
terminate a walk once it returns.  To accomplish this, we  compute 
$\mathbb{P}(R_{i,k})$ by taking \emph{modified} powers of the adjacency matrix 
$\boldsymbol{A}$.  Define $zd(\boldsymbol{A})$ as a function that sets the diagonal entries of 
$\boldsymbol{A}$ to 0. Then we compute the modified $k$-th power of $\boldsymbol{A}$ as:
\begin{equation}
\label{Ak}
\boldsymbol{A}^{(k)} = 
\begin{cases}
    \boldsymbol{A} & \text{if } k = 1, \\
    zd(\boldsymbol{A}^{(k-1)})\boldsymbol{A} & \text{otherwise}
\end{cases}
\end{equation}
where $\boldsymbol{A}$ is the original adjacency matrix. Note that we use $\boldsymbol{A}^{(k)}$
instead of $\boldsymbol{A}^{k}$ to distinguish our modified matrix multiplication
from ordinary matrix multiplication.  To understand the purpose
of setting the diagonal to 0, consider that the expression $\boldsymbol{A}^{k}\boldsymbol{A}$
extends the $k$-step walks of $\boldsymbol{A}^{k}$ with the 1-step walks of $\boldsymbol{A}$.
Setting the diagonal entries of $\boldsymbol{A}^{k}$ to 0 causes walks that
return to the origin at step $k$ to terminate at the origin, which
satisfies the definition of a self-absorbing walk.  It is easy
to see analogies among the terms of~\autoref{PRet} and~\autoref{Ak} --
that is, between $1 - \sum_{x=1}^{k-1}\mathbb{P}(R_{i,x})$ and $zd(\boldsymbol{A}^{(k-1)})$
on the left, and $\mathbb{P}(Next_{i,k})$ and $\boldsymbol{A}$ on the right.

Computing $\boldsymbol{A}^{(k)}$ gives us the values of $ReturnSteps_{i,k}$ and
$Steps_{i,k}$.  Since $a^{(k)}_{i,i}$ is
$ReturnSteps_{i,k}$ and $\displaystyle\sum_{j}a^{(k)}_{ij}$ is
$Steps_{i,k}$, we redefine $\mathbb{P}(Next_{i,k})$ as follows:
\begin{equation}
\label{PNext2}
\mathbb{P}(Next_{i,k}) = 
\begin{cases}
    \frac {a^{(k)}_{ii}} {\displaystyle\sum_{j}a^{(k)}_{ij}} & \text{if } \displaystyle\sum_{j}a^{(k)}_{ij} > 0 \\
    0 & \text{if } \displaystyle\sum_{j}a^{(k)}_{ij} = 0 
\end{cases}
\end{equation}

Thus instead of taking increasing powers of an adjacency matrix,
and counting and dividing at each step, we take increasing powers of
a transition probability matrix $\boldsymbol{P}$.  Specifically, we design the 
following procedure for 
computing the distribution of expected return probabilities for each
vertex $i \in V$ and for steps $1, \ldots, k$:

\begin{enumerate}
\item Initialization
\begin{enumerate}
\item Initialize $x$ to 1.
\item Initialize $k$ to the number of steps to perform.
\item Initialize the transition probability matrix $\boldsymbol{P}$.
\end{enumerate}
\item Iteration
\begin{enumerate}
\item If $x = k$, terminate.
\item Compute $\boldsymbol{P}^{(x)}$ with \autoref{Ak}.
\item Read the values from the diagonal of $\boldsymbol{P}^{(x)}$; the value
$\boldsymbol{P}^{(x)}_{i,i}$ is the expected return probability for node $i$ at
step $x$.
\item Increase $x$ by 1.
\end{enumerate}
\end{enumerate}

The complexity of this algorithm, computed naively, is $O(kn^{3})$.
It can be computed much more efficiently using sparse matrices.  
Arithmetic operations on them are proportional to $nnz$, the number 
of non-zero entries.  However, as $k$ increases, $\boldsymbol{P}^{(k)}$ becomes 
less sparse and the benefits of sparse matrix multiplication decrease. 
Let $nnz_k$ be the number of non-zero entries in the $k$-th matrix 
computed by our algorithm.  Then with sparse matrices the time 
complexity of our algorithm is $O(k \times nnz_k)$.  We present another
optimization in \autoref{Power measures} when we discuss return 
probability as a measure of power.

The return probability for an entire network and some $k$ can be computed
by averaging the return probabilities of all nodes:
\begin{equation}
\label{PGlobal}
\mathbb{P}(R_k) = \frac{1}{n} \sum_{i \in V} \mathbb{P}(R_{i,k})
\end{equation}

This network-wide measure can be used in the same fashion as the
node-specific form of the measure.  It can generate a distribution
of probabilities, either step-wise or cumulative.  For easy comparison
to other measures, it can be reduced to a scalar value by taking
the step-wise or cumulative return probability at a given $k$.  As
expected, \autoref{PGlobal} reaches its highest value in a dyad,
where it is 1, regardless of $k$.  Any other network has a network
return probability $<1$.

\subsection{Validation}

To validate that our method correctly computes expected
return probabilities, we conduct an experiment similar to those in
\cite{Borgatti2005}.  In this case we release a random walker on a
100-node scale-free network and count the number of times it returns or
fails to return to the origin for walks of particular lengths.

\begin{figure}
\centering
\subfloat[]{\includegraphics[height=3in,clip]{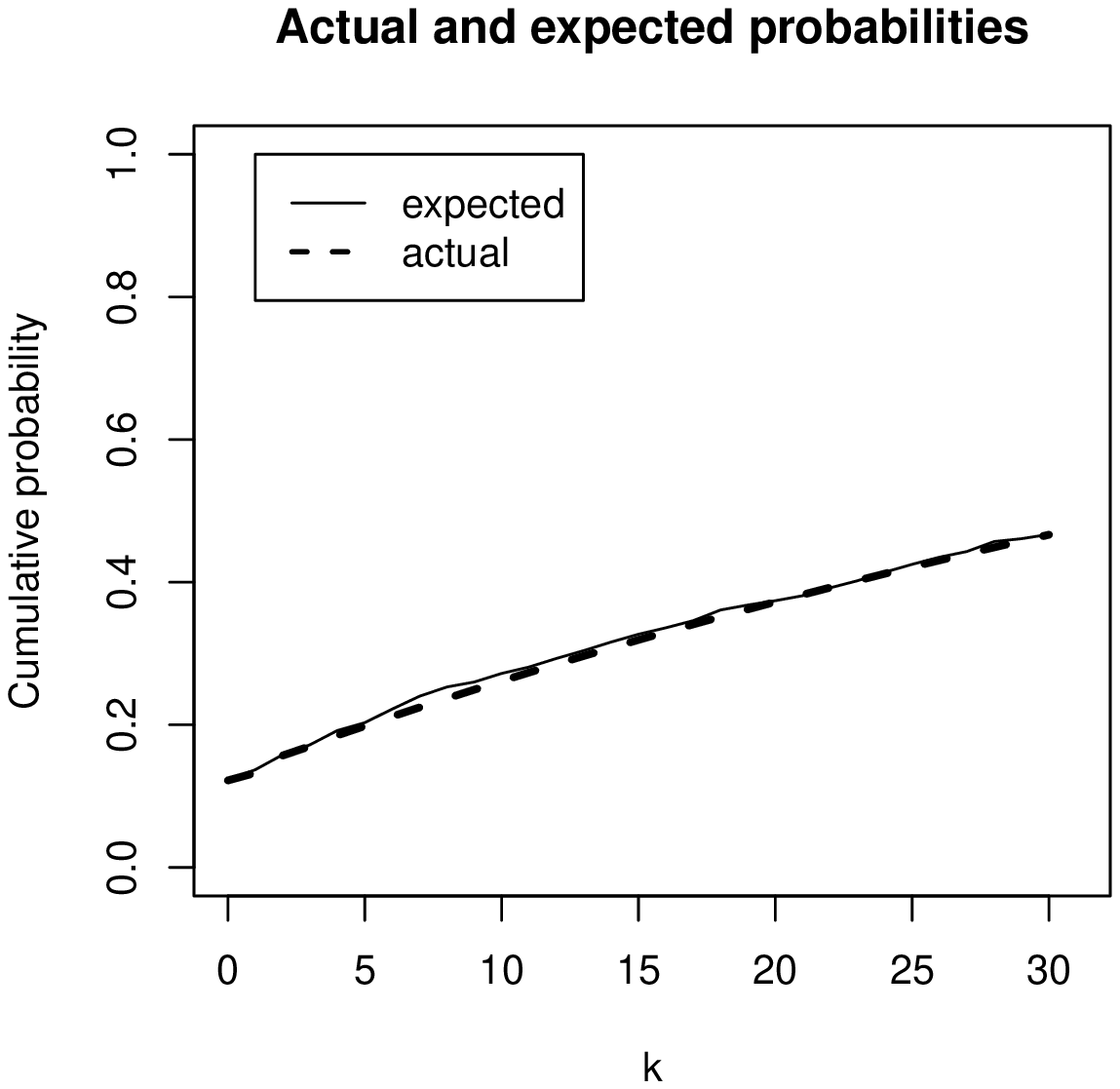}} \\
\subfloat[]{\includegraphics[height=3in,clip]{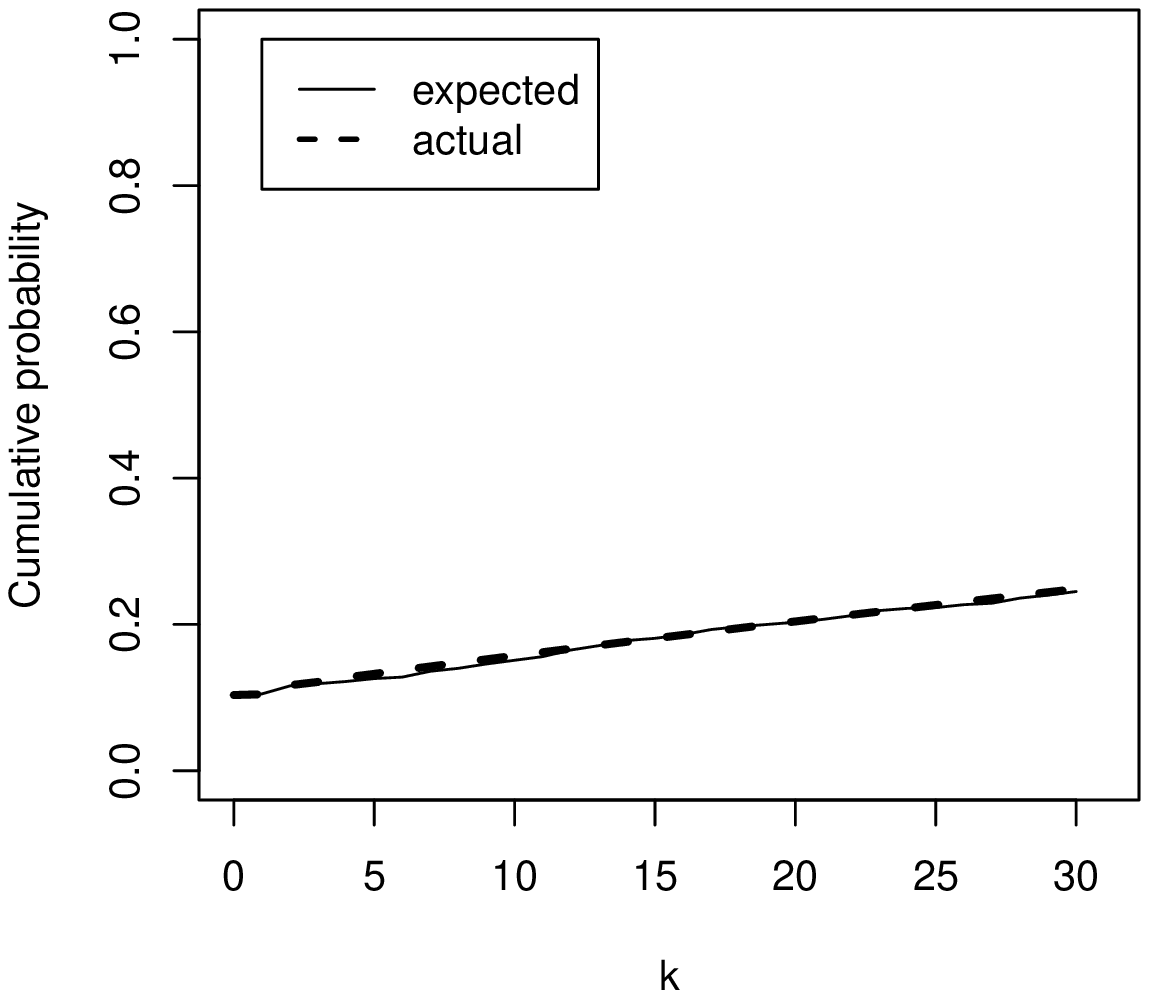}}
\caption{Expected and actual return probabilities for walks of increasing
length from two different nodes in a 100-node scale-free network.  Actual return
rates were averaged over 1000 walks started at each node.}
\label{fig:expected-and-actual}
\end{figure}

Since the walks are random, we do not know the length of any given
walk in advance.  Rather, we start walking from a node $i$ and if we
return to $i$ in $k$ steps, we record this fact and start another walk.
When we have completed some number of walks using a node $i$ as the
origin, we compute the actual return rates for $i$ on walks of
length $k$ by counting the number of times we returned to $i$
on a walk of length $k$ and dividing by the total number of walks.
As shown in~\autoref{fig:expected-and-actual}, for two different nodes 
in a scale-free network, expected return probabilities
computed by our algorithm match well with the actual return rates 
found after 1000 walks, for different $k$ values. 

\begin{table}
\centering
\begin{tabular}{|c|c|c|}
    \hline
Number of walks & Complete graph & Small-world network \\
    \hline
4 & 0.87 & 0.89 \\
6 & 0.90 &  0.91 \\
8 & 0.95 &  0.95 \\
10 & 0.98 & 0.78 \\
12 & 0.91 & 0.82 \\
14 & 0.92 & 0.94 \\
16 & 0.96 & 0.93 \\
18 & 0.97 & 0.98 \\
20 & 0.97 & 0.93 \\
22 & 0.97 & 0.98 \\
24 & 0.99 & 0.95 \\
26 & 0.99 & 0.99 \\
28 & 0.99 & 0.95 \\
30 & 0.99 & 0.99 \\
32 & 1.00 & 0.99 \\
34 & 1.00 & 0.95 \\
36 & 0.99 & 0.98 \\
38 & 0.99 & 0.99 \\
    \hline
\end{tabular}
\caption{High correlation between actual and expected returns in simulations
with increasing number of walks on two 100-node graphs.}
\label{table:corr-with-increasing-walks}
\end{table}

In general our experience is that the number of walks required in order
to achieve a high correlation between expected and actual returns is
small, as shown in~\autoref{table:corr-with-increasing-walks}.
We have also tested with other graphs of varying sizes, which exhibit 
the same general correspondence between expected return probabilities 
and actual return rates.

\section{Return Probability and Other Measures} \label{Return Probability and Other Measures}

If return probability is a meaningful measure, in what sense is it so?  
We answer this question by exploring relationships between
return probability and two measures of power -- beta centrality and the
graph-theoretical power index (GPI) -- and subgraph centrality.  We
find that return probability resembles a measure of exclusionary
power when $k=2$, which we call the \emph{P\'{o}lya power index} (PPI).
We also find that return probability is equivalent to subgraph centrality when
$k>2$.  It also has asymptotically significant running-time advantages
over both beta centrality and subgraph centrality.

\begin{figure}
\centering
\subfloat[5-person network (3 positions)]{\includegraphics[scale=0.5]{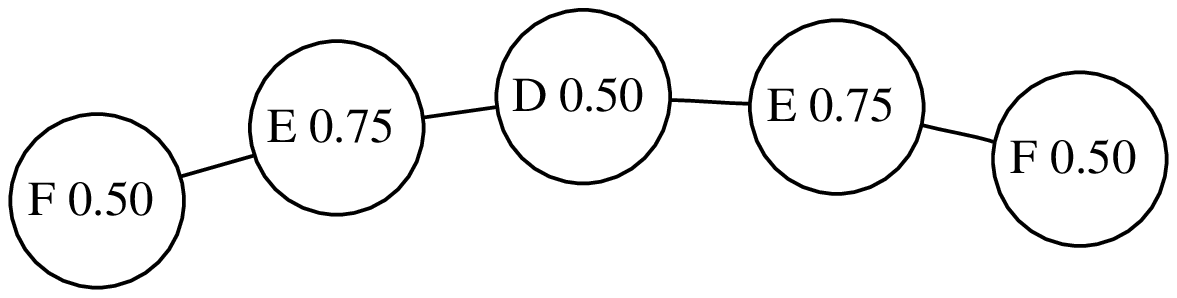}}
\subfloat[7-person network (3 positions)]{\includegraphics[scale=0.5]{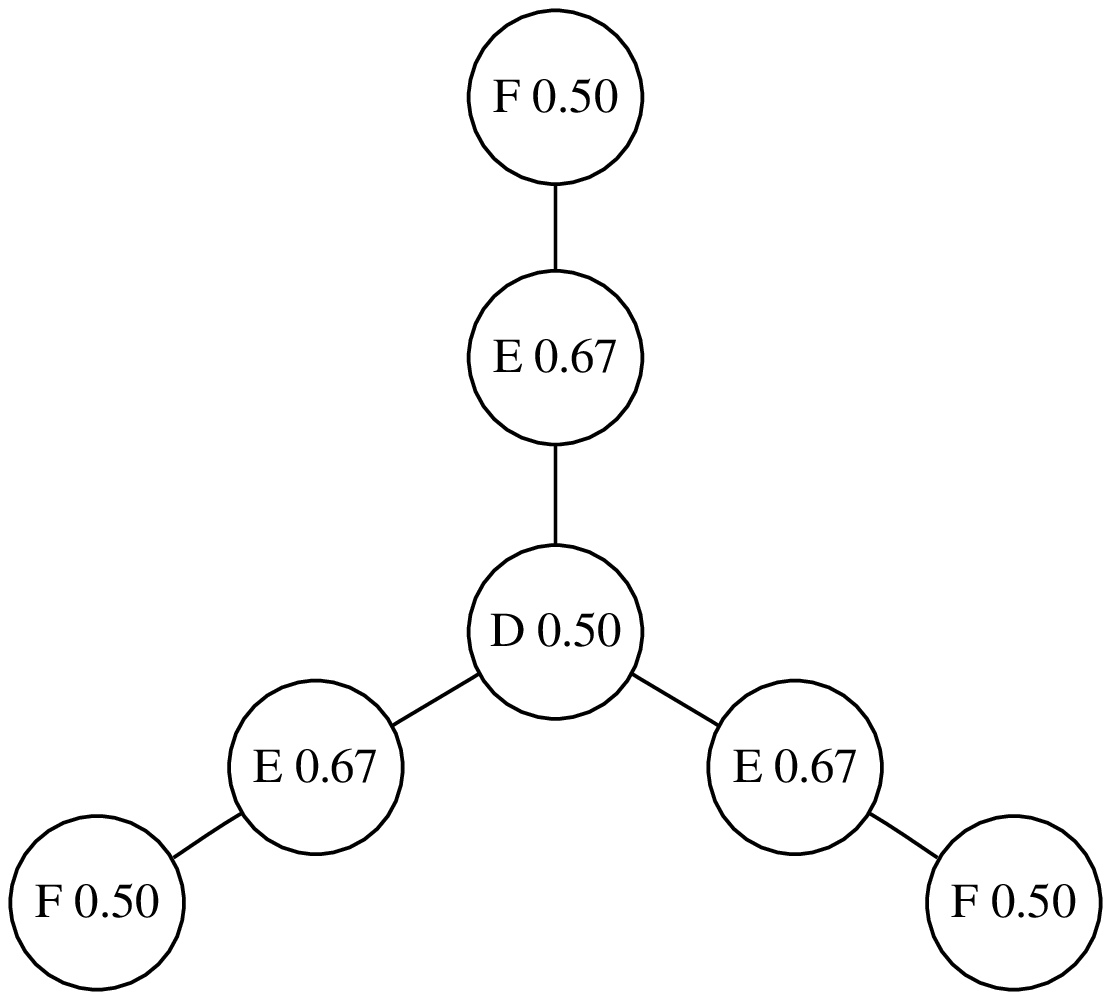}} \\
\subfloat[10-person network (3 positions)]{\includegraphics[scale=0.45]{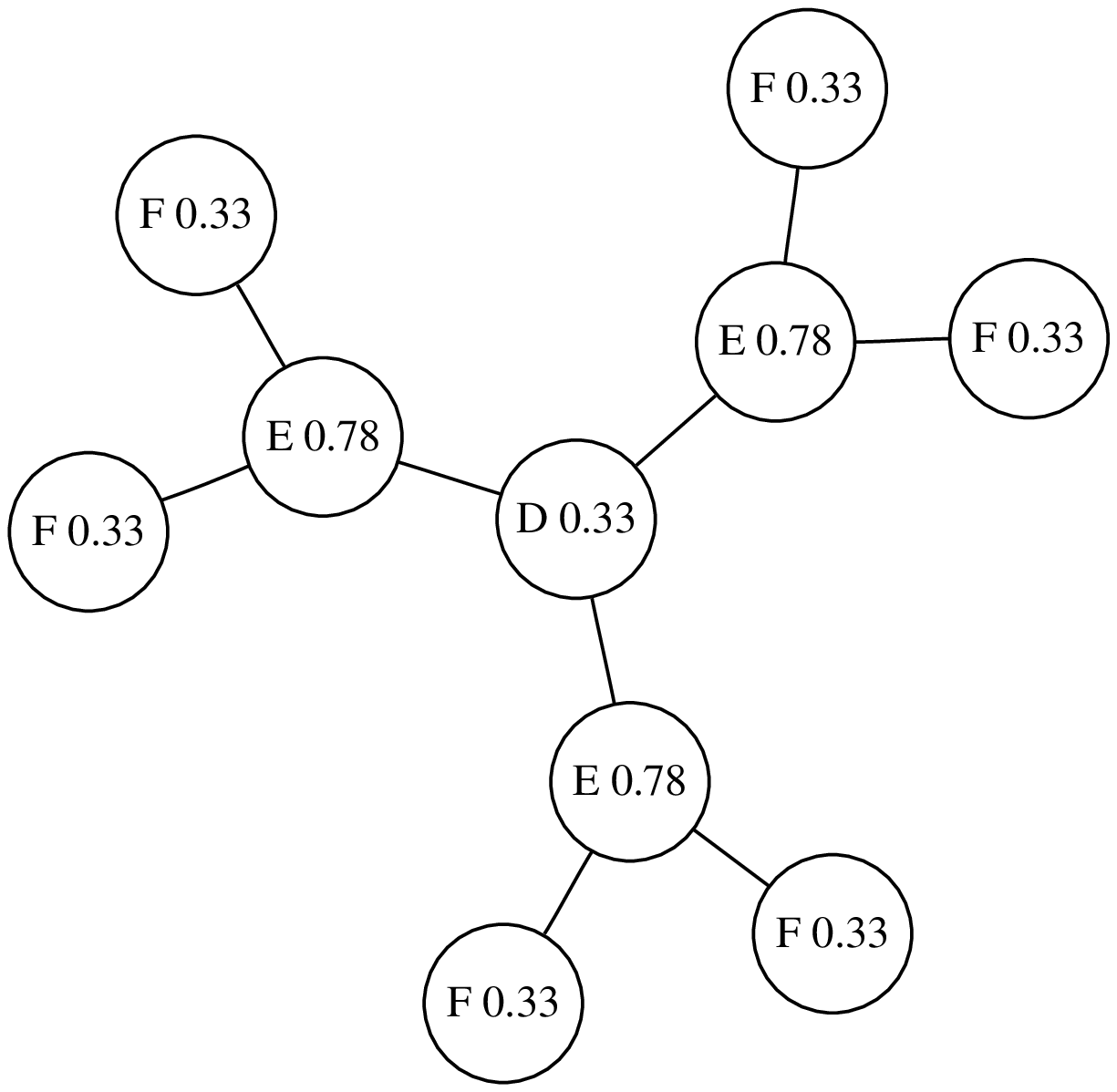}}
\subfloat[13-person network (3 positions)]{\includegraphics[scale=0.45]{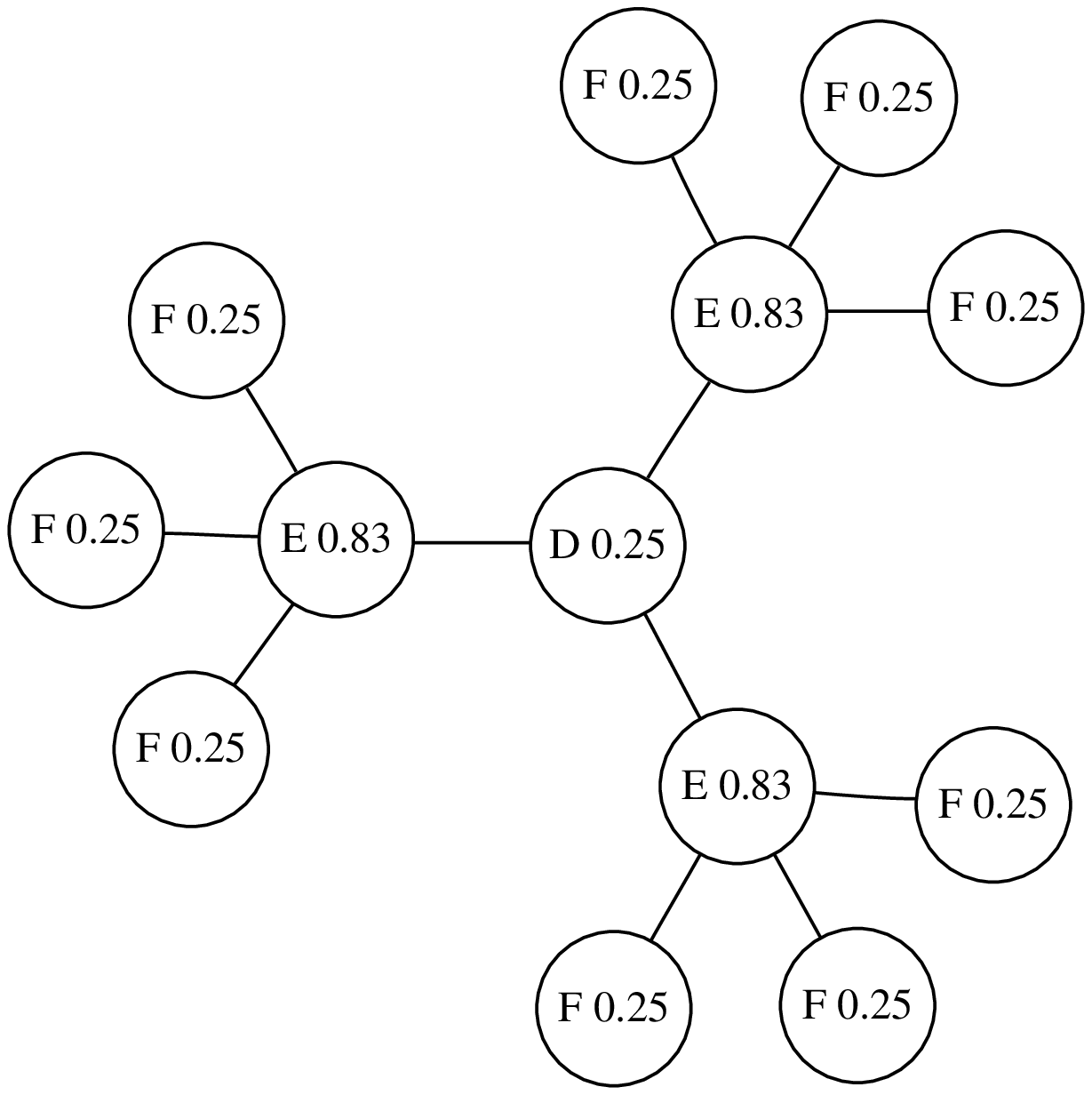}}
\caption{Strong-power exchange networks illustrating the Power-Dependence
Theory (PDT) experiments of Cook~\emph{et al}.  People with structurally
similar positions in the network are assigned the same category
($E,D,F$).  People are labeled with their category and their 2-step
return probability.  According to PDT, power
is distributed in these networks according to the relation $E>D=F$.
The 2-step return probabilities agree with the predictions of Cook 
\emph{et al}. (See also Figure 1 in \cite{Cook1978}.)}
\label{fig:cook-graphs}
\end{figure}

\subsection{Power Measures}
\label{Power measures}

Beta centrality and the GPI originated in the competing theories
of exchange networks, thus many of the experiments conducted with
them are concerned primarily with acts of exchange.  However, they
-- like return probability -- may be appropriate for identifying
powerful nodes in non-exchange networks as well.  When we refer to
these measures, including return probability, as measures of power,
we mean power in the broadest sense of the term, not just limited
to exchange networks.  Thus, while we rely on results in the exchange
network literature to illustrate the relationships among return
probability, beta centrality, and the GPI, we do not think of return
probability necessarily as a mechanism for generating predictions
for the outcomes of network exchange experiments.  We subscribe to
the distinction between ``power as a potential and power as an
activity'' (\cite{Willer1992}) and claim only that return probability
can identify nodes in powerful positions.

\begin{figure}
\centering
\subfloat[L4]{\includegraphics[scale=0.45]{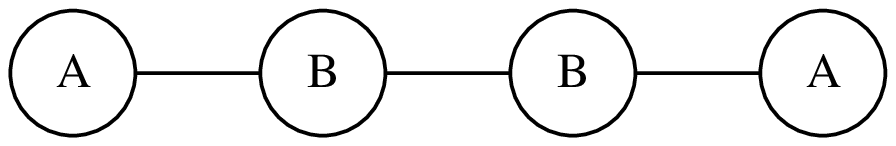}} 
\subfloat[Stem]{\includegraphics[scale=0.45]{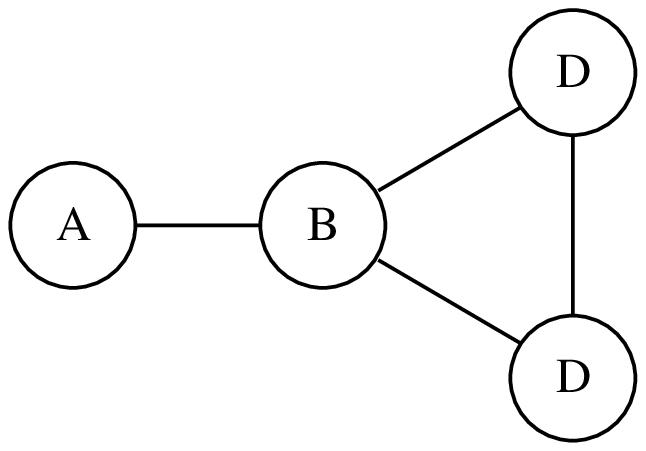}}  \\
\subfloat[L5-Stem]{\includegraphics[scale=0.45]{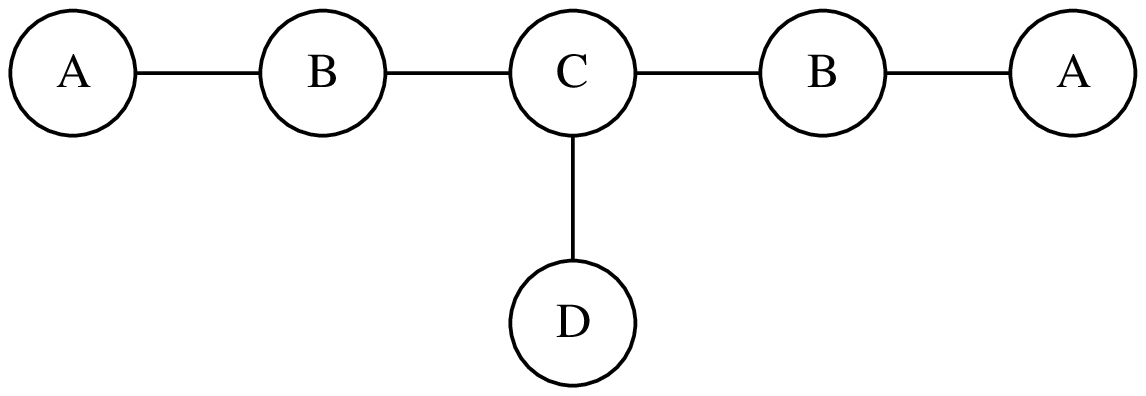}} \\
\subfloat[K-Stem]{\includegraphics[scale=0.45]{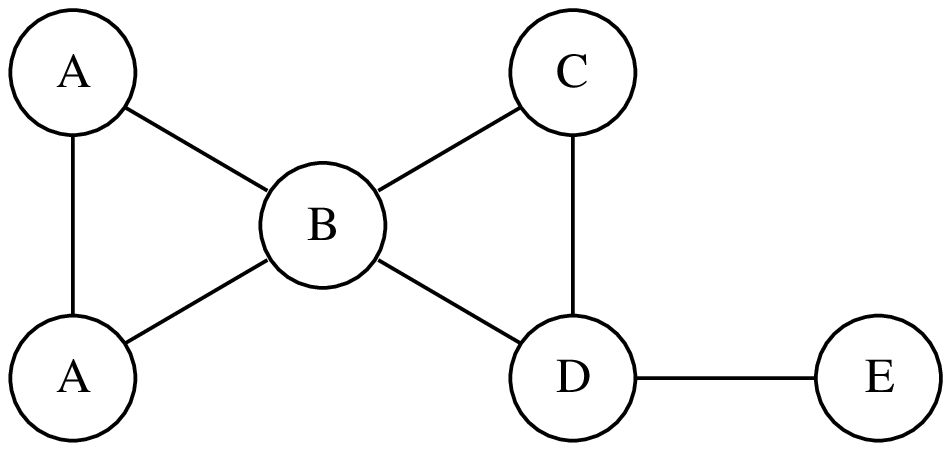}}  \\
\subfloat[Borg-6]{\includegraphics[scale=0.45]{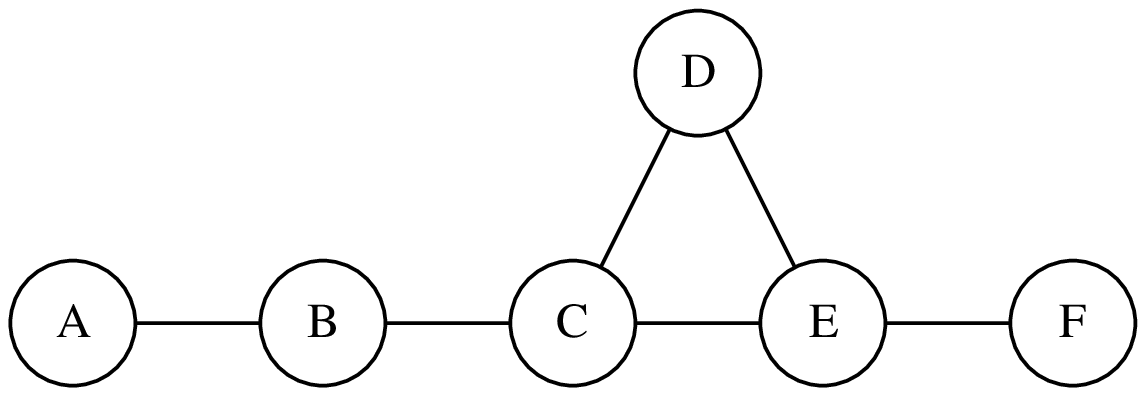}}
\caption{Weak-power networks.}
\label{fig:weak-power-networks}
\end{figure}

We begin with the results of an early experiment in exchange network
theory~\cite{Cook1978}, using them to compare return probability and
beta centrality only.  The networks used in this early experiment, shown in \autoref{fig:cook-graphs}, were
discovered later to be strong-power 
networks~\cite{1993Markovsky}.  We then turn to the weak-power networks
of \autoref{fig:weak-power-networks} and use them to compare return
probability, beta centrality, and the GPI.

Beta centrality is motivated in part by the fact that in \cite{Cook1978},
classical centrality measures -- degree, betweenness, and closeness
-- failed to predict the outcomes of experiments with
negatively-connected exchange networks.
In an exchange network, actors exchange objects
of value.  An exchange network is connected positively or negatively.
Imagine an exchange network consisting of three participants $A$, $B$,
and $C$; $A$ is connected to $B$, $B$ is connected to $C$.  If the
network is positively connected, an act of exchange between $A$ and $B$
does not preclude a concurrent act of exchange between $B$ and $C$.
If the network is negatively connected, $B$ cannot exchange with $A$ and
$C$ at the same time.  Here we are primarily concerned with
negatively-connected networks.

The GPI originated in the network exchange literature as well.  It is
associated with Elementary Theory, a competitor
to PDT that has itself fared quite well in the
experimental literature.  There are several versions of the GPI.  Here we
rely on Markovsky's version and results from the \emph{Social Networks}
special issue on exchange networks~\cite{Markovsky1992}.  That version
is known to produce contradictory results under some conditions (see
footnote 2 of \cite{Willer2008}), but to our knowledge those conditions
do not apply to these particular results.  Many of the revisions of the
GPI that occurred after the formulation of the original GPI -- including
Markovsky's -- make use of a probability of a node being excluded from
exchange.

\begin{table}[ht!]
\centering
\begin{tabular}{|c|c|c c c c c c|}
    \hline
Network  & Measure & \multicolumn{6}{c|}{Edge} \\
    \hline
L4 &           & A-B & & & & & \\
  & $Return\ probability$       & $<$ & & & & & \\
  & $Beta\ centrality$       & $<$ & & & & & \\
  & $GPI$      & $<$ & & & & & \\
    \hline
L5-Stem &           & A-B & B-C & C-D & & &  \\
  & $Return\ probability$      & $<$ & $=$ & $>$ & & &  \\
  & $Beta\ centrality$      & $<$ & $\boldsymbol{\ll}$ & $>$ & & &  \\
  & $GPI$     & $<$ & $=$ & $>$ & & &  \\
    \hline
Stem &           & A-B & B-D & & & &  \\
  & $Return\ probability$      & $<$ & $>$ & & & &  \\
  & $Beta\ centrality$      & $<$ & $>$ & & & &  \\
  & $GPI$     & $<$ & $>$ & & & &  \\
    \hline
K-Stem &           & A-B & B-C & B-D & C-D & D-E &  \\
  & $Return\ probability$      & $<$ & $>$ & $<$ & $<$ & $>$ &  \\
  & $Beta\ centrality$      & $<$ & $>$ & $\boldsymbol{\gg}$ & $<$ & $>$ &  \\
  & $GPI$     & $<$ & $>$ & $<$ & $<$ & $>$ &  \\
    \hline
Borg-6 &           & A-B & B-C & C-D & C-E & D-E & E-F \\
  & $Return\ probability$      & $<$ & $>$ & $>$ & $<$ & $<$ & $>$ \\
  & $Beta\ centrality$      & $<$ & $\boldsymbol{\ll}$ & $>$ & $<$ & $<$ & $>$ \\
  & $GPI$     & $<$ & $>$ & $>$ & $<$ & $<$ & $>$ \\
    \hline
\end{tabular}
\caption{Relative power of nodes in weak-power networks according to
return probability, beta centrality, and Markovsky's GPI.}
\label{table:retprob-beta-gpi}
\end{table}

We first show that 2-step return probability does not contradict with some
PDT predictions.  The people in the networks in
\autoref{fig:cook-graphs} are labeled with a category determined by a
person's position in the network and the person's cumulative 2-step
return probability.  The networks are isomorphic to four exchange
networks analyzed in both \cite{Cook1978} and \cite{Bonacich1987}.
For simplicity of presentation, the figures only include the more
profitable solid lines from the original network; the dashed lines
are excluded.  Following both Cook \emph{et al}, and Bonacich, we
compute return probability using only the solid lines.  Basing their
hypotheses on PDT, Cook \emph{et al} predicted
that the power distributed among the actors in these networks would
reach the equilibrium $E > D=F$.  Their prediction was supported by 
both a laboratory experiment and computer simulations.  The fact that
2-step return probability matches the predicted equilibrium for all
graphs exactly is shown in the labels of \autoref{fig:cook-graphs}.
One obtains the same results using beta centrality, 
with one exception: in the 7-person network, the relation remains $D >
E > F$, the same as a conventional centrality.

It was discovered by Markovsky \emph{et al} that there are in fact different 
classes of networks~\cite{1993Markovsky}.  The networks used
in the aforementioned experiment are strong-power networks -- networks
in which the relations of exchange are stable and the nodes in positions
of relative strength dominate their exchange partners.  There are also
equal power networks in which no actor has an advantage.  A third class --
weak-power networks -- are structurally somewhere between strong and
equal power networks.  \autoref{fig:weak-power-networks} contains several
examples of weak-power networks.  The discovery of the different classes
of networks brought a deeper understanding of the nature of the networks
themselves.  In strong-power networks, for example, the actors are
clearly divided between those with high power and those with low power;
low-power actors are only connected to high-power actors.  Such networks
are bipartite or very close to bipartite.  We revisit bipartivity
in \autoref{Subgraph centrality}.

\autoref{table:retprob-beta-gpi} shows the relative power for all
connected nodes in the networks in \autoref{fig:weak-power-networks}
computed by return probability, beta centrality, and
the GPI\footnote{To compute beta centrality, we use UCINET 6 for
Windows~\cite{citeulike:1030331} with $\beta=-0.2$.   For GPI, we
use the values of $GPI_3$ and $p$ whenever $GPI_3 = 1$ in Markovsky's
Table 1~\cite{Markovsky1992}.}
If the power of a node $A$ exceeds the
power of neighbor $B$, then the value of edge $AB$ is listed as $>$.
Here we consider the GPI to be a touchstone.  Where beta centrality
or return probability disagree with the GPI, the symbol is doubled and
in bold (e.g., $\boldsymbol{\ll}$).  This abuses notation somewhat
but is readable enough.  Note that return probability agrees with the
$GPI$ for all edges of all networks.  The only disagreements are between
the GPI and beta centrality.  In L5-Stem, beta centrality 
identifies $C$ as the most powerful node in the network,
whereas the GPI and 2-step return probability have $B$ on equal footing
with $C$.  In K-Stem, the GPI and return probability compute that $B$
is less powerful than $D$, and beta centrality holds the opposite.
There is a similar disagreement over the edge $BC$ in Borg-6.  Generally, it
seems that in these particular weak-power networks, beta centrality
has difficulty identifying the power conferred on a node $i$ when it
is connected to a node $j$ that has no other exchange opportunities.
In such a configuration, $i$ is always guaranteed the option of trading
with a relatively powerless neighbor.

Since in this case we only need to concern ourselves with $k=2$,
we can reformulate the original algorithm from \autoref{Algorithm}
to be even more parsimonious.  The doesn't reduce the time complexity
over our original algorithm in a meaningful way, because the number
of non-zero entries in a sparse matrix is already related to the
number of vertices and edges.  However, it provides a form of the
equation that can easily be computed when the network is represented
in memory as a graph not as a matrix.  In honor of George P\'{o}lya,
this is the \emph{P\'{o}lya power index} ($PPI$):

\begin{equation}
\label{PolyaPowerIndex}
PPI = \mathbb{P}(R_{i,2}) = \frac{1}{deg(i)}\sum_{j: (i,j) \in E}\frac{1}{deg(j)}
\end{equation}

When applying this computation to an entire graph, each edge appears
twice in a summation, so it runs in time $O(n+m)$.  This is
significantly faster than beta centrality.  Exact implementations
of beta centrality require a somewhat costly $O(n^3)$ matrix
inversion.  Approximate implementations sum the first $k$ terms of
beta centrality's infinite series; just like return probability,
the approximate version of beta centrality can be computed with
sparse matrices, which makes the matrix multiplications sub-cubic.
Even then our algorithm is faster.

\begin{figure}
\centering
\includegraphics[width=4in]{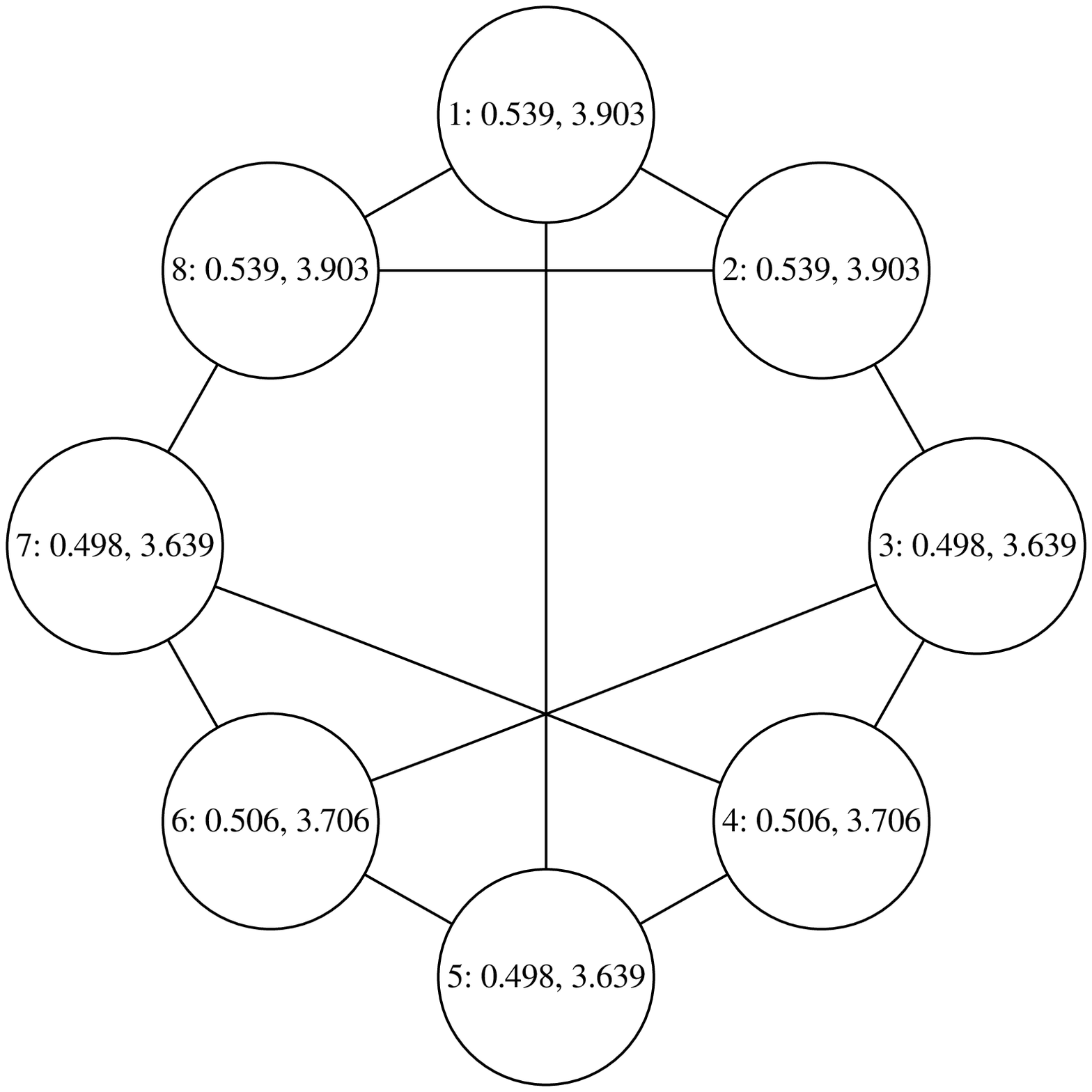}
\caption{A regular graph in which degree, closeness, betweenness,
and eigenvector centrality are the same for all nodes, but return
probability and subgraph centrality vary according to
each node's position in the graph.  Vertices are labeled with the node
number, its return probability for $k=5$, and its subgraph centrality.
Return probability identifies the same sets of nodes ($\{1,2,8\}$,
$\{3,5,7\}$, $\{4,6\}$) as subgraph centrality.}
\label{fig:retprob-and-sg-cent}
\end{figure}

\subsection{Subgraph Centrality}
\label{Subgraph centrality}

Return probability can also be used to identify interesting nodes when
$k>2$.  Both \cite{Estrada2005} and \cite{Bonacich2007} have noted that
in some regular graphs, eigenvector centrality \cite{Bonacich1972}
is equivalent to degree centrality.  Both subgraph centrality and
return probability are able
to distinguish nodes from one another in such graphs.  The label of
each node in \autoref{fig:retprob-and-sg-cent} shows cumulative return
probability for $k=5$ and subgraph centrality.  Both measures identify
the same groupings of nodes and thus have the same discriminatory power.
This is no surprise, because both measures are expressed as diagonals
of powers of some matrix representation of a graph.  Additionalluy,
both measures count trivial closed walks (i.e. a closed walk made
from a path starting at node $i$ and the return path to $i$ along
the same edges).  However, return probability counts only \emph{unique}
trivial and non-trivial closed walks -- which it accomplishes by
way of self-absorbing walks -- whereas subgraph centrality counts
\emph{all} trivial and non-trivial closed walks.

\begin{figure}
\centering
\includegraphics[scale=0.7]{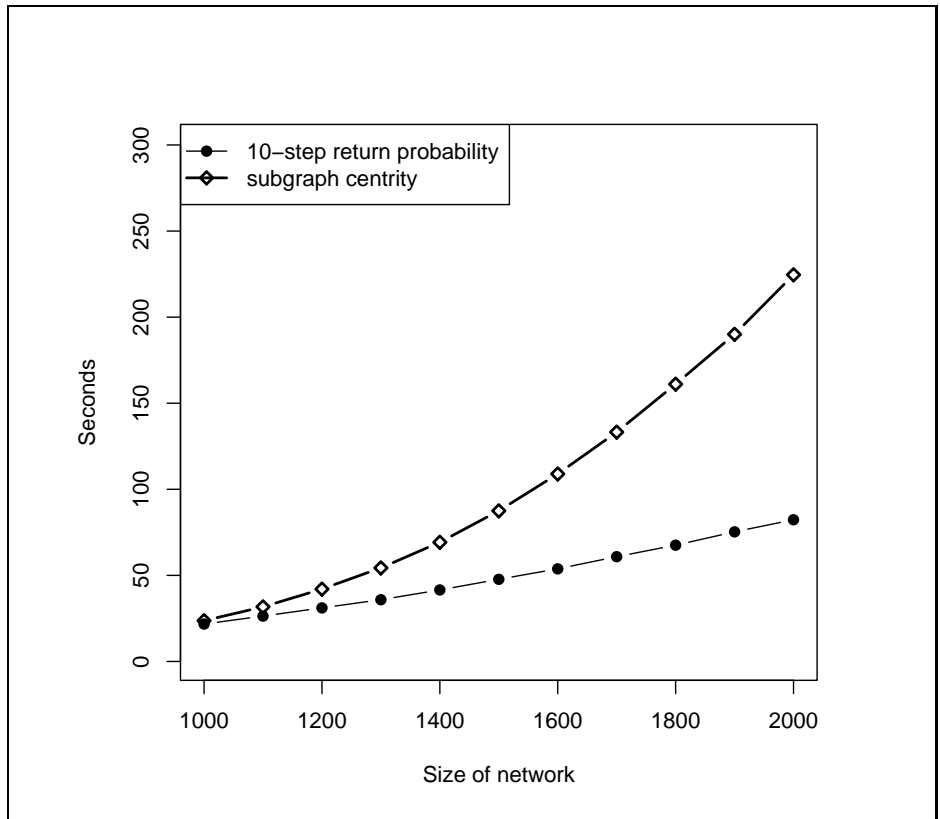}
\caption{Running time for 10-step return probability and subgraph
centrality in small-world networks ranging from 1000 to 2000 nodes.}
\label{fig:retprob-and-sgcent}
\end{figure}

\begin{figure}
\centering
\includegraphics[width=4in]{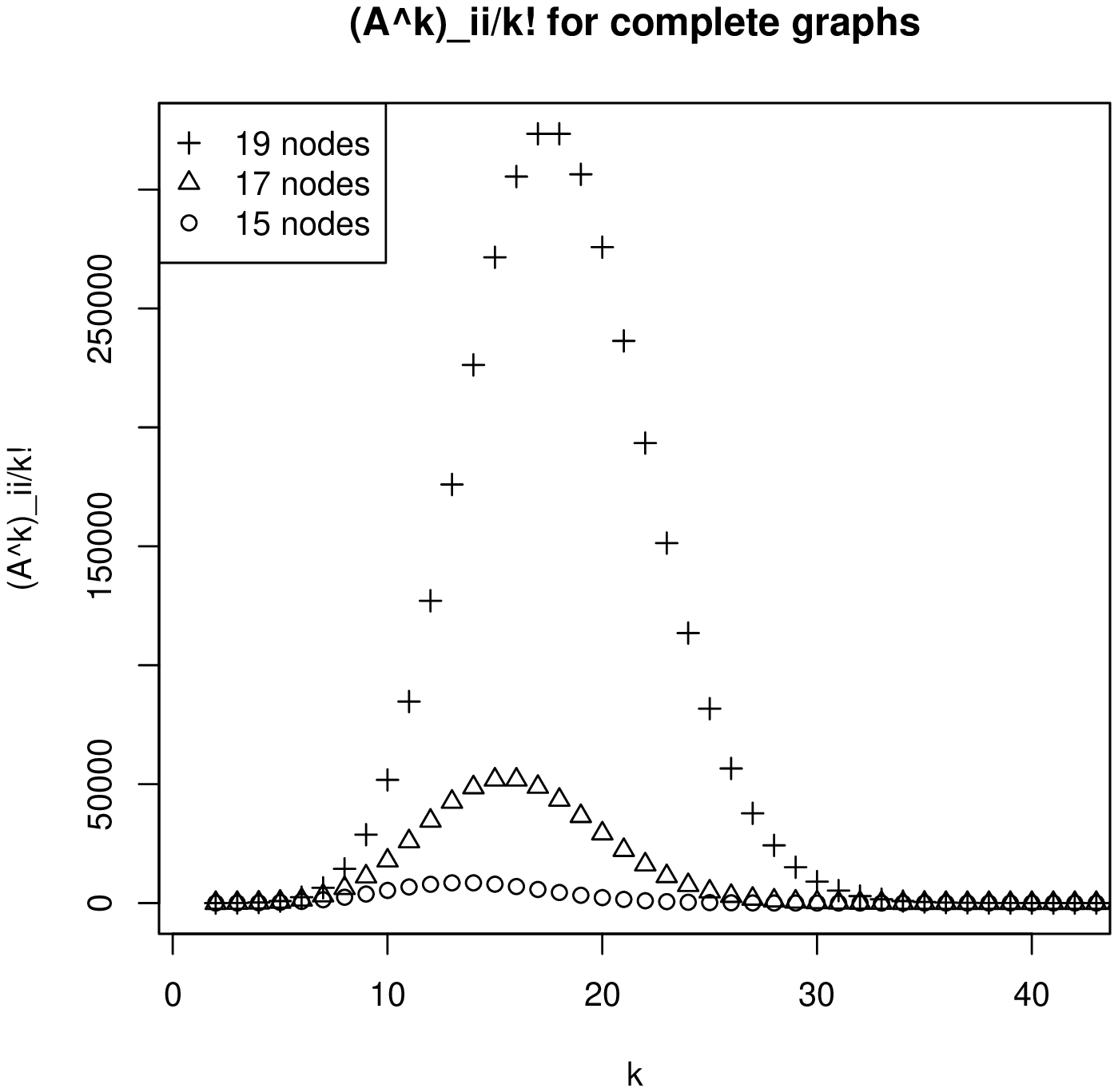}
\caption{The number of closed walks of length $k$ counted by subgraph
centrality for complete graphs of different size.}
\label{fig:sg-ratio}
\end{figure}

For small motifs, return probability runs more quickly than subgraph
centrality regardless of the size of the network, albeit at the cost of
greater memory consumption due to sparse matrices not being multiplied in
place.  This can be seen in \autoref{fig:retprob-and-sgcent}, which shows
elapsed time for 10-step return probability and subgraph centrality in
small-world networks of varying size.  The benchmarking program is written
in the Python programming language and makes use of the SciPy library for
scientific computing \cite{scipy}.  The eigendecomposition function is
\texttt{scipy.linalg.eigh}, which in turn uses the robust LAPACK and BLAS
 linear algebra libraries~\cite{lapack}.  We ran the program on a
computer with a 1.8 Ghz AMD Athlon 2200 processor and 768 MB RAM.
Given that matrix multiplication is highly parallel, a distributed
MapReduce-style system is an appropriate solution to the problem of
computing return probability for networks too large to fit into the
memory of a single computer,

In a complete graph, both the number of closed walks of length $k$
and the factorial of $k$ grow quickly.  Ultimately, for sufficiently large 
$k$, $k!$ reduces ${\boldsymbol{A}^k}_{ii}/k!$ to 0.  However, the \emph{rate of
growth} of ${\boldsymbol{A}^k}_{ii}$ also increases rapidly with the order of $\boldsymbol{A}$.
In \autoref{fig:sg-ratio}, the value of $k$ at which ${\boldsymbol{A}^k}_{ii}/k!$
reaches its maximum increases as the number of nodes increases.
Thus, the lengths of the closed walks counted by subgraph centrality are
sensitive to the size of the network.  By contrast, return probability
allows one to select exactly the lengths of the walks considered by
the measure.

\begin{figure}
\centering
\includegraphics[width=4in]{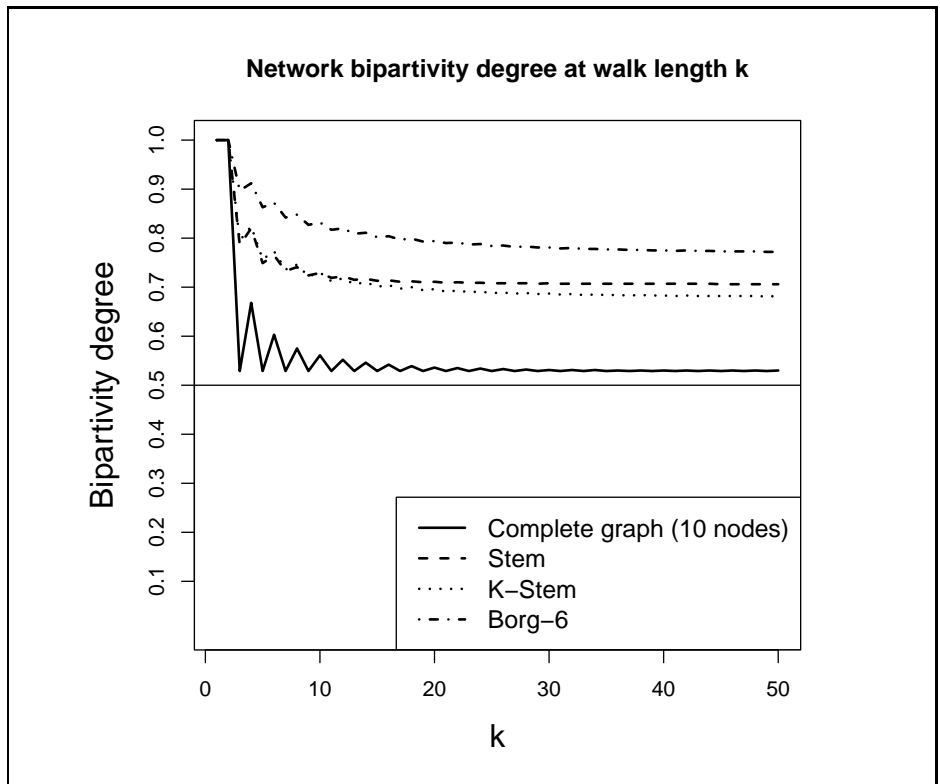}
\caption{The network bipartivity degree of non-bipartite weak-power
networks for walks of increasing length.}
\label{fig:bipartivity-xchgnets}
\end{figure}

\begin{figure}
\centering
\subfloat[]{\label{fig:bipartivity-hybrid}\includegraphics[height=3.25in,width=3.25in]{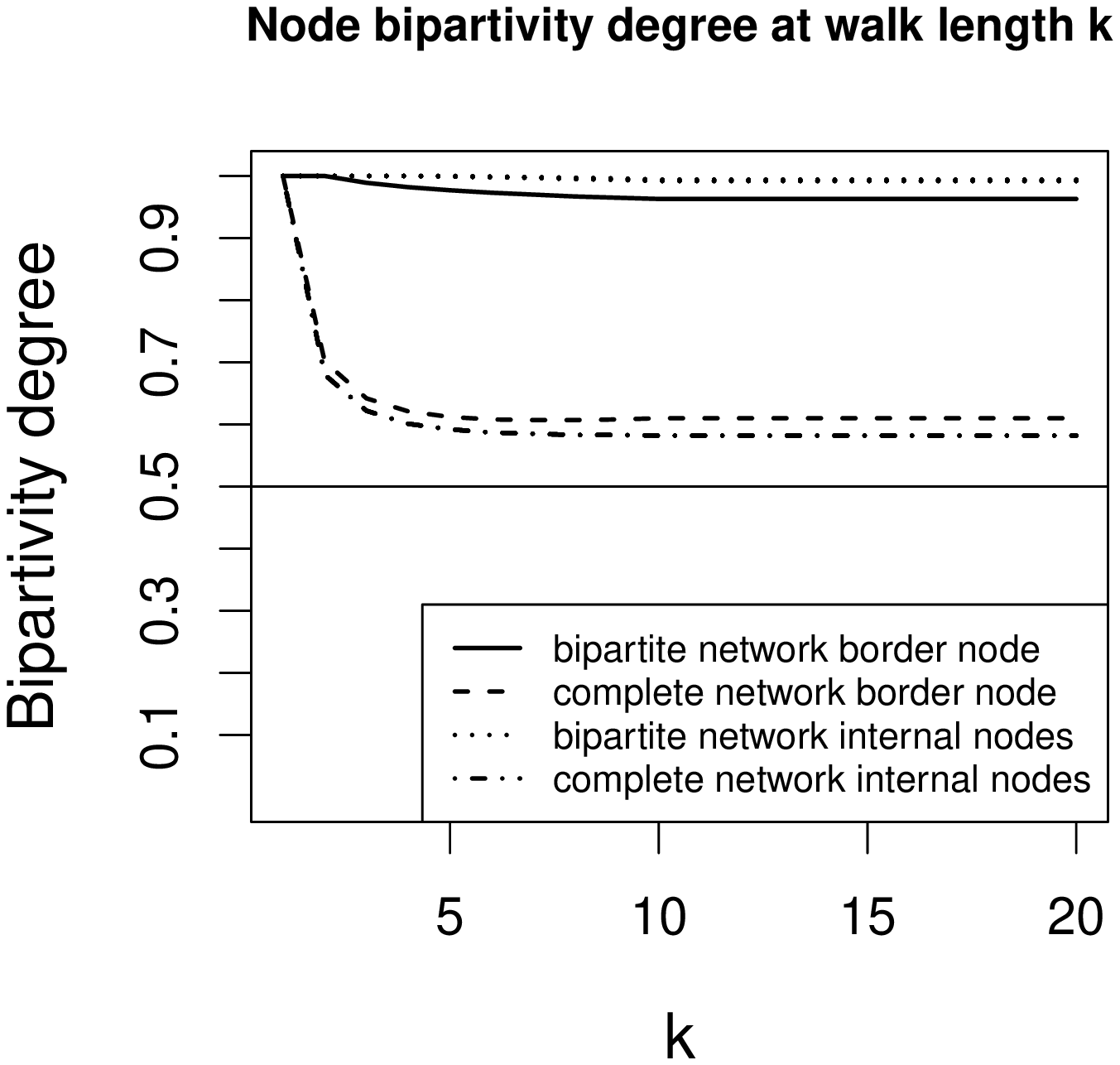}} \\
\subfloat[]{\label{fig:bipartivity-hybrid-zoom}\includegraphics[height=3.25in,width=3.25in]{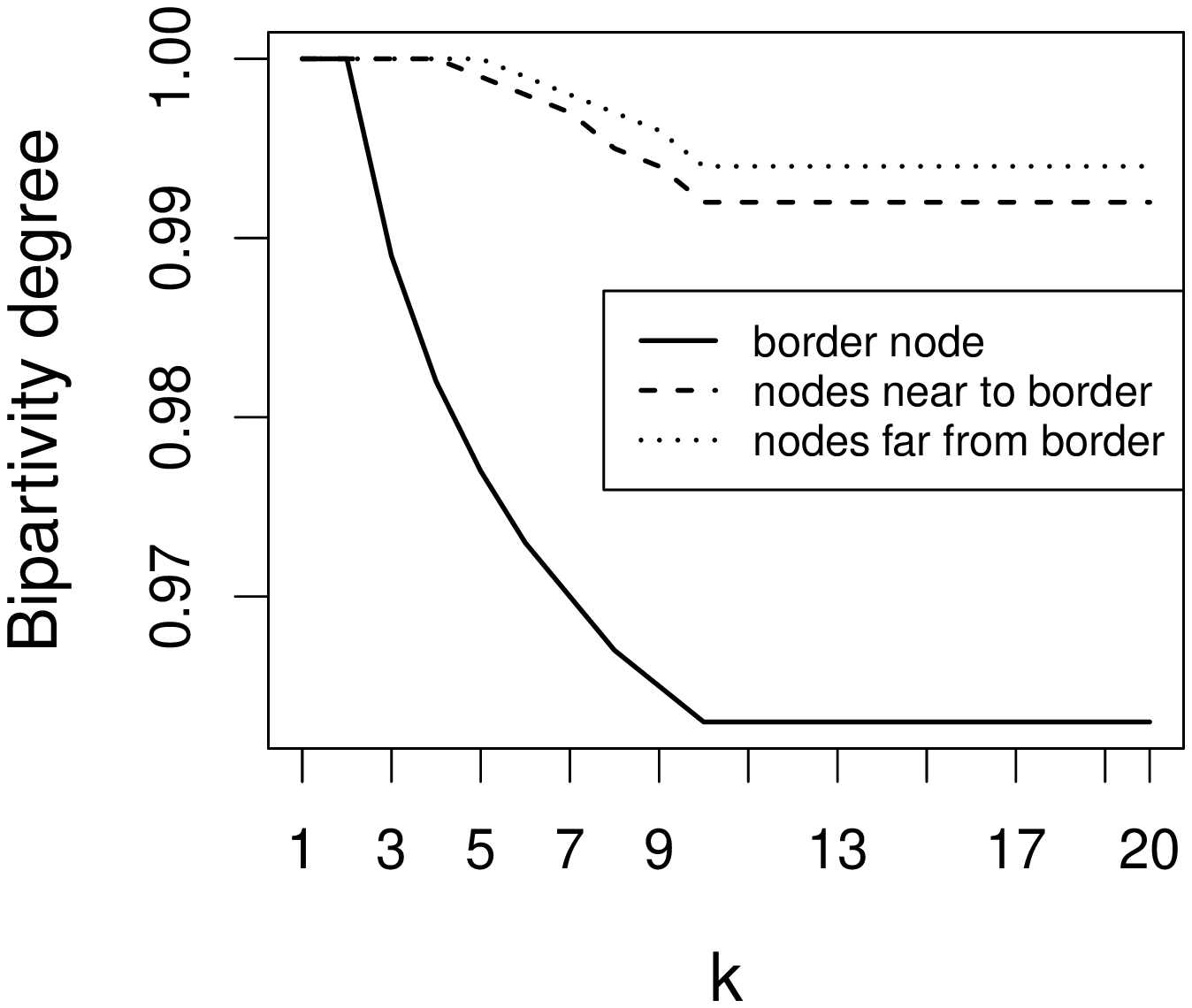}}
\caption{The node bipartivity degree of nodes in a network that
consists of a complete bipartite network joined by one edge with a
complete graph.  Note in (b) that a node's bipartivity degree remains
1 longer depending on its proximity to the border node.}
\end{figure}

Both subgraph centrality and return probability can be used to
quantify \emph{bipartivity degree}, a measure of how close a network
is to being bipartite~\cite{PhysRevE.68.056107}.  Since a bipartite
network contains no odd cycles, the number of even cycles
divided by the number of cycles is 1.  When computed for an entire
graph, subgraph centrality is expressed as \cite{PhysRevE.72.046105}:
\begin{equation*}
SC = \frac{1}{n} \sum_{i=1}^{n} SC_i = \frac{1}{n} \sum_{i=1}^{n} e^{{\lambda}_i}
\end{equation*}
Which leads to the following equation for a network's subgraph
centrality~\cite{PhysRevE.72.046105}:
\begin{equation*}
\beta(G) = \frac{SC_{even}}{SC} = \frac{\sum_{j=1}^{n} cosh \lambda_j}{\sum_{j=1}^{n} e^{\lambda_j}}
\end{equation*}

The same can be computed in terms of a network's cumulative return
probability for walks up to length $k$:
\begin{equation*}
\beta(G_k) = \frac{\mathbb{P}(R_k)_{even}}{\mathbb{P}(R_k)} = \frac{\sum_{j: (j \equiv 0\text{ mod }{2})}^{k} \mathbb{P}(R_j)}{\sum_{j=1}^k \mathbb{P}(R_j)}
\end{equation*}
Or of a node $i$'s cumulative return probability:
\begin{equation*}
\beta(G_{i,k}) = \frac{\mathbb{P}(R_{i,k})_{even}}{\mathbb{P}(R_{i,k})} = \frac{\sum_{j: (j \equiv 0\text{ mod }{2})}^{k} \mathbb{P}(R_{i,j})}{\sum_{j=1}^k \mathbb{P}(R_{i,j})}
\end{equation*}

Bipartivity degree is typically a value in the range [0.5, 1].  For
a bipartite network -- such as the strong-power networks in
\autoref{fig:cook-graphs} -- $\beta(G)$ is 1.  When even and odd
closed walks contribute equally $\beta(G)$ is 0.5; the bipartivity
degree of a complete graph approaches 0.5 as both $k$ and the size
of the graph grow.  \autoref{fig:bipartivity-xchgnets} shows that
the bipartivity degree of the non-bipartite weak-power networks
differs even as $k$ increases.

Clearly it is possible to use bipartivity degree to distinguish
communities that tend towards homophily from communities that tend
towards heterophily.  \autoref{fig:bipartivity-hybrid} shows the
node bipartivity degree for all nodes in a graph that consists of
a complete bipartite graph and a complete graph joined by a single
edge $e$.  The ``border'' nodes are the nodes made adjacent by $e$.
Not only are the nodes in the two different networks clearly
distinguishable from each other, but the border nodes show a clear
divergence away from the bipartivity degree of the other nodes in
their cohort and towards each other.  Narrowing in on the nodes in
the bipartite graph, we can see in \autoref{fig:bipartivity-hybrid-zoom}
that the bipartivity degree of the nodes farthest from the border
node remains 1 for longer than that of the nodes in the same set
as the border node.

\section{Discussion} \label{Discussion}

For some measures, the $k$ in ``$k$-step'' is a parameter of the
measure itself -- for example, the $k$ parameter of return probability.
When $k=1$, return probability is the inverse of degree.  As $k$
increases, the walks touch a larger neighborhood of nodes.  Thus if one
wishes to compute return probability for a neighborhood of a specific
size, one simply chooses the appropriate value of $k$.  The $\beta$
parameter of beta centrality also suggests the size of the neighborhood
around node $i$ that is included in the computation.  When $\beta$ is 0,
beta centrality is akin to degree.  When the absolute value of $\beta$
is small, only proximal neighbors are considered, and the neighborhood
grows as $|\beta|$ increases.  However, $\beta$ itself does not
indicate the exactl length of walks.  However, a cut-off point can
be clearly defined when computing the approximation of beta centrality,
because then one can sum the first $k$ terms of the series
and terminate.  The GPI also has a parameter, $e$, but it
is unrelated to the scope of the computation; the GPI is computed from
one up to the diameter of the network.  Subgraph centrality has no walk
length parameter, but as we describe in \autoref{Subgraph centrality},
the size of the neighborhood it considers varies, being determined only
by size of the network and the factorial
denominator.

Both beta centrality and subgraph centrality, being expressed as
infinite sums of open or closed walks, have to cope with the problem
of convergence.  The root of the problem is that walks never
terminate.  In an undirected graph, the number of walks of length
$k$ is always greater than the number of walks of length $k-1$.
They deal with this by weighing walks inversely by length (beta
centrality) or by factorial of length (subgraph centrality).  When
computing beta centrality or subgraph centrality, the weighting is
the same regardless of the structure of the network.  Return
probability also converges, but for different reasons, and weighting
is determined by the underlying structure of the network itself.
In the star graph $S_{1,n}$, 2-step return probability for the
center node is 1, and for the others it is $1/n$.  Take $S_{1,4}$.
Since it is bipartite, there are no odd-length paths, and the return
probabilities for steps $1, \ldots, 6$ are $0.0, 0.25, 0.0, 0.1875, 0.0,
0.1407$.  \autoref{PCond} automatically scales step $k$ by the
complement of step $k-1$ so that it is a portion of what remains.
Without \autoref{PCond}, they are $0.0, 0.25, 0.0, 0.25, 0.0, 0.25$,
so \autoref{PNext} is $0.25$ at every even-numbered step.  This is
obvious when you consider that if the walk has not returned to the
origin, it is always at the center of the star.  \autoref{PCond}
scales the return probability of step 4 down by the portion of
probability already accounted for by step 2.  So if the 2-step
return probability for a node $i$ increases due to an edge being
added to one of $i$'s neighbors, the return probability for $i$ at
all subsequent steps is reduced in proportion to the increase at
step 2.

Beta centrality and the GPI measure power by having odd-length paths
contribute positively and even-length paths contribute negatively to
the value the measure computes for a node $i$.  The 1-paths contribute
positively to $i$'s power, just as a node with higher degree has
higher degree centrality.  The 2-paths detract from $i$'s power because
they provide neighbors with the opportunity to exchange with some node
other than $i$.  Two-step return probability functions similarly in that
it increases when the degree of $i$'s neighbors decreases.  The less the
opportunity $i$'s neighbors have to exchange with a node other than $i$,
the greater is $i$'s 2-step return probability and the greater its power.

Using ``$k$-step measure'' as the name for the measures discussed
in this paper makes the category somewhat more general, albeit in
name only.  The category could be made even more general.  $K$-step
measures compute some value for a node $i$ by considering a sequence
of increasingly larger sets of nodes by starting with a set containing
only $i$, then adding the neighbors of $i$, and the neighbors of
those neighbors, and so on.  In mathematical morpholgy, dilation
$\delta(g)$ is an operation on a subgraph $g$ of graph $G$ which
adds to $g$ the nodes of $G$ adjacent to those of $g$
\cite{2007AdPhy..56..167C}.  A $d$-dilation is the application of
$\delta$ $d$ times:
\begin{equation*}
\delta_d(g) = \underbrace{\delta(\delta(\ldots(g)\ldots))}_d
\end{equation*}
The process of constructing the sequence of sets considered by a
$k$-step measure is just a series of $k$ dilations.  This process
is a constrained form of dilation in so far as it always starts
with a subgraph of \emph{one} node.  A more general definition --
one that permits inclusion of a greater number of measures -- would have two
components: (1) a generic process that constructs a sequence of
node sets by repeatedly dilating an \emph{arbitrary} subgraph $g$ of $G$;
and (2) an unspecified computation which takes as input the sequence
of node sets generated by the first component.  

It may be that 2-step return probability is a close approximation
of structural exclusive power because it resembles the model of
Markovsky's \emph{WeakNet} simulation software in which actors ``(i)
seek  one  exchange  per round,  (ii) seek  to exchange  with  a
randomly  selected  other,  and  (iii) keep  seeking exchange  in
a  given  round  until  no  more  potential exchange partners
remain'' \cite{Markovsky1992}.  If $A$ and $B$ are connected, the
probability that $A$ seeks to exchange with $B$ and $B$ seeks to
exchange with $A$ is just the product of the inverse of the degrees
of $A$ and $B$.  Similarly, 2-step return probability is the
probability that $A$ will be chosen as an exchange partner by
\emph{any} randomly-selected neighbor.

The performance improvements that return probability shows over
beta centrality and subgraph centrality should be understood in the
context of practical analysis.  If one wishes to compute subgraph
centrality and one already has the eigenvalues and eigenvectors of
the adjacency matrix -- say, for use by some other algorithm --
computing return probability would be wasteful.  An analyst who
needs to identify powerful nodes in a somewhat large network may
be able to get an answer in a reasonable amount of time by using the
approximate version of beta centrality algorithm and stopping the
computation after a small number of walks.  Where we imagine return
probability being most useful is in the analysis of extremely large
networks.  When a network's nodes number in the millions (e.g.
online social networks, the World Wide Web), the
computational efficiency of return probability makes it an attractive
alternative to beta centrality or subgraph centrality.

\section{Conclusion} \label{Conclusion}

We have presented an algorithm for computing return probability for
networks.  The measure is probabilistic, so it requires no
normalization, and it permits exact control over the size $k$ of
the neighborhood it forms around a node.
Because it shows agreement with
beta centrality and the GPI, the \emph{P\'{o}lya
power index} appears to
be a measure of relative power in networks.  It is also just as
capable as subgraph centrality at classifying nodes based on features
of a network that can only be identified by looking at longer-distance
relationships.  Further, the time complexity of return probability
-- $O(n+m)$ for the \emph{P\'{o}lya power index} and proportional to $k \times nnz_k$
when $k>2$ -- is less than either beta centrality or subgraph
centrality and lends itself to easier analysis of extremely large
networks.

\bibliographystyle{h-elsevier}
\bibliography{retprob}{}

\end{document}